# Touching the Sky: The Use of Arduino in Transferring Telescopic Light to Haptic Vibrations

Wanda Diaz-Merced [1], Ruoning Lan

- [1] Astroparticle and cosmology laboratory: wanda.diaz.merced@gmail.com

- [2] Brown University; ruoning_lan@brown.edu

Abstract: Astronomers often only "look at" mathematically purified data sets at risk of buffering and/or ignoring visually ambiguous critical points from the original information. The chaotic nature of outer space and the limitation of visual displays command for much more than visual display of information and for the integration of other sensorial modalities during data exploration. Haptic real time devices may enrich the detection of astronomical events that otherwise would escape the eyes. Departing from the Harvard Astronomy Lab and Clay Telescope's Orchestar (color Arduino), we present the work in progress of the Proof of Concept (PoC) of a sensitive yet simple device to Bluetooth transfer real time color into haptic motion built by Adafruit components. We assemble 2 Adafruit nRF52840 feather express with the RGB color sensor and haptic driver respectively to trigger vibrations according to the color variation from external light sources. For the rigor of astronomical studies, we aim to transmit as more data in the shortest amount of delay time possible. In addition, a transparent hexagon cover will be mounted on the color sensor to maximize absorbed light from the telescope. The device aims to be a "translator" for people to "see" "hear" and "feel" the hidden information from the original data set. We also present its application in the calculation of complex astrophysics quantities such as the masses of solar coronal mass ejections.

Keywords: sonification; haptic; Arduino; real-time software; accessibility; haptic; coronal mass ejection

1. Introduction

For centuries, astronomy has remained primarily visual. With data automatically converted to images and graphs, astronomers decode the sky through pinpointing critical turns, tracing variation tendencies, and calculating statistical values. This has become the default way to examine the universe, even though mathematical techniques impose buffering, linearization, standardizations, leakages and other setbacks [1-2] added to the limitations imposed by nature on human visual perception [3-4]. There could be supplementing means to extract science. The capability of audition or touch to aid vision in astronomy suggests a broader perspective and possibilities of novel insights driven by new devices that transform the astronomy information, (in our PoC case light visible or invisible), into sounds and haptic vibration [5].

Computer purified dataset sometimes omits important points hidden in noises from the chaotic outer space environment, especially for those distant yet essential objects such as extragalactic nebulas and black holes. In addition to providing alternatives to data representation, by applying sounds and haptic vibrations into mathematically complex dataset, visually ambiguous information could be more easily detected. Cardiff University's Black Hole Hunter is an example of using sonification to uncover the signal of black holes buried in cumulative noises as it travels through the space [6]. Application such as xSonify [7], LHC sound

[8], the university of Berkeley [9] and others has been the pioneer to allow users to input data for mainstream research purposes and transform it to differing pitches.

In this report we present our progress on the use of Bluetooth technology to transfer real time the light detected from the Orchestar [10] to haptic vibration. Our aim is to make the haptic response, real time, scientifically accurate, while at the same time making a device simple to build and accessible to larger bandwidth of audiences of a variety of backgrounds (professional astronomers, amateur astronomy, individuals with disabilities). With the Orchestar feather express connected to an optical telescope, a user holding the haptic Bluefruit feather express as far as 5 to 7 meters away from the telescope will be able to feel the real-time haptic feedback corresponding to the color detected by the feather express Arduino at the telescope.

Symmetrically this progress report of our PoC we will discuss how Bluetooth BLE connection is used in communication between two microcontrollers, the reduced minimum delay time for data transfer, and the planned sensor experiments to find the appropriate waveforms of vibration.

On section 2 the reader will find the materials employed, section 2.1 the methods employed and on section 3 the status of the PoC project at the moment of submission.

2. Materials and Methods

The original device, color sensor Arduino made by Harvard [10], was constructed using Adafruit components for its compatibility, implementation, and transportability of the original coding, usage of libraries and accessible documentation. We decided to use the Adafruit feather express nRF5240, Adafruit haptic controller DRV2605 due to its libraries providing for the usage of different waveforms to be indicative of different colors.

As shown in Figure 1 and Figure 2, we construct two separate Arduinos. In Figure 1, the Color Arduino is assembled according to the Harvard Orchestar to transfer light into sound. In Figure 2, the Motor Arduino, centered around the feather express, consists of a motor controller DRV2605, and a motor disk. We aimed to connect the two feather expresses with Bluetooth, and then transfer the sound data obtained from the Color Arduino to the Motor Arduino, in order to process it into haptic vibration.

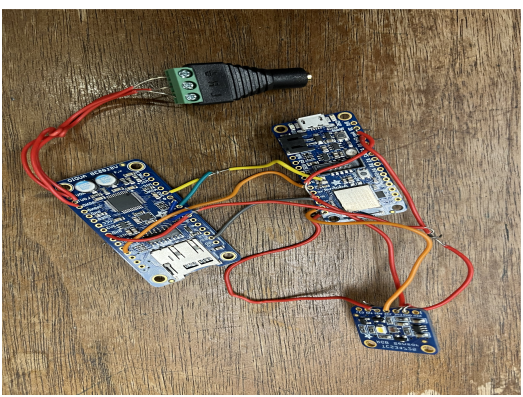
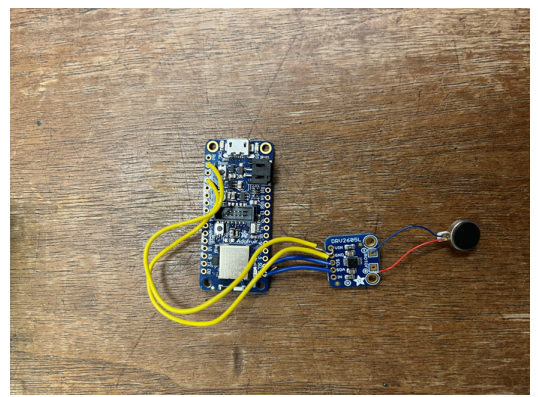

Figure 1. Color Arduino.                              Figure 2. Motor Arduino

In Figure 3 and figure 4 the wiring diagrams are shown for people to build their own Arduinos.

Setting up the Bluetooth to recognize and exchange information was the most essential yet arduous part. We imply Adafruit and Nordic to simplify the process so that interested peoples of all backgrounds may take advantage of this Bluetooth technology.

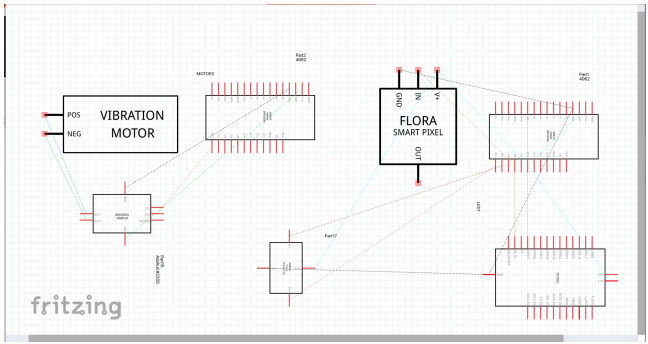

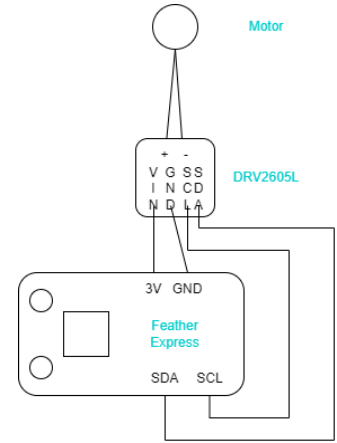

Figure 3 Color Arduino wiring

Figure 4.  Motor Arduino wiring

First step was to check for the stability of the Bluetooth connection. We initiated that by establishing a connection with the built-in Bluefruit connection [11-12] on an iPhone and android. Based on the motion of celestial objects and the instrument tracking time, once we successfully connected the Adafruit feather express to the telephone, we established as a test for the the (connection only) to be stable for 20 minutes without interruption at proximities of 5 to 7 meters between the Arduino and the mobile phone.

We proceeded to set up the Universal Unique Identifier or UUID's. The pairing and setting of the UUIDs are extremely difficult as the full information is not readily at hand. The only examples we could find from Adafruit at the time were the heart rate monitor [13], though it did not provide links to the characteristic and service values corresponding to the feather express.

We then tested the connection between the two Bluetooth Arduinos for 20 minutes nonstop at distances of 5 to 7 meters. The testing was performed in rooms without other Bluetooth devices, thus minimizing external disruption. A message would be printed on the serial monitor once the central and peripheral are connected or interrupted. The connection was manually detained after 30 minutes by closing the serial monitor. The serial monitor connection is extremely important as it displays real-time light intensity acquired by the Orchestar [10]. We took advantage of the peripheral payload of 20 milliseconds to ensure data resolution, defining the peripheral as the data acquiring Orchestar and the central as the haptic motor Arduino.

3. Results

In order to transfer information from one Arduino to another, we will have to define Bluefruit dual role [14].

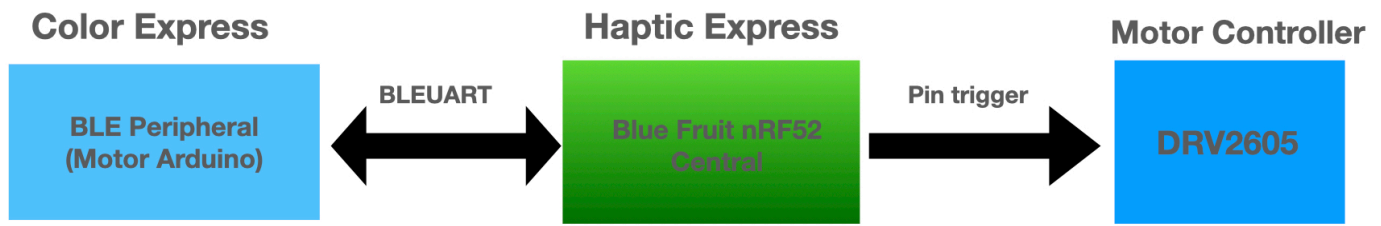

Figure 5. Flowchart of Bluefruit central and peripheral role [14].

We define the haptic Arduino as the peripheral via UART, which means it can send messages to the central Haptic Express which triggers a the motor controller. In this case, the haptic Arduino receives color RBG information from the color Arduino, processes it, and then sends it to the motor controller to let it vibrates.

The Bluetooth connection was established stably. The haptic Arduino beacons Bluetooth signals to look for its peripherals (color Arduino), indicated by the beeping blue lights. Once it finds the connection, the indicator light will stay blue, therefore opening the data pave way. At the same time, if the colour express is connected to the computer the monitor displays the current RGB reading every fraction of a second, updating the continuous light input. [ref hyman]

For our current progress, the next step is to devise a function relating color hue to vibration waveform. In order to corresponds light to haptic, we need a mathematical function to find the best vibration frequency that's sensitive while not numbing people. Subsequent sensorial experiments are expected to find such frequency.

To let the color Arduino absorb as more light as possible, we also plan to add a hemisphere on top of the color sensor mimicking those of infrared sensors [15].

The connection of the devices was tested on a telescope. The Orchestar connected to the telescope using its original case and holding the haptic motor at distances of a diameter of 5 and 7 meters from the telescope. The moon is a very bright object with magnitude of approximately -12. The device was effective triggering its haptic vibration inside the delay time defined by the Adafruit components, but as the moon does not change intensity, it is too early to admit its efficiency triggering haptic response that are due to lower light intensities.

The Adafruit haptic motor has a very extensive library that we hope will facilitate to define different haptic responses and intensities on different color and intensity of light.

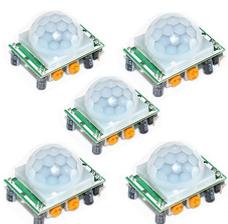Figure 6. Infrared sensor [16].

Because of the cover's hexagonal structure, this infrared sensor could direct more light to the sensor and take in more light than without it. We were inspired by the design to add a similar cover on top of our RGB color sensor, thus furthering the rigor of astronomical studies. In that way, the color sensor could respond to larger amount and greater distance of light, adding to the accuracy of our device.

With celestial objects moving rapidly in the range of the telescopes, it is imperative to shorten the delay time for the value sampled by the color sensor to change, be transferred to the haptic Arduino and trigger the vibration feedback. The advantage of Adafruit is that it already has a minimized delay time of 20 milliseconds between wired Arduinos. Starting from that, we need to devise an efficient algorithm to shorten the processing time and therefore achieve a delay less than 50 milliseconds.

4. Discussion

During assembling, we originally used a breadboard to connect pieces instead of soldering. Albeit convenient to build, the breadboard built-ons are not suitable for carryon or compact purposes. So, we chose to wire the Arduinos directly through the built-in pins. As a result, our device could be compacted inside a carrier box to receive signal while the colour Arduino is mounted on the telescope.

The device has to be experimented on the best packing for the haptic Arduino. The packing has to be sensorially tested as not to buffer any vibrations triggered by dim light intensities. It is also important to consider that this device has to be usable by the widest range of people. Then portability becomes also a primary aim.

Equally the haptic responses have to be tested on the scanning of images. For those purposes we created a coronal mass ejection (CME) lesson to allow us to scan a known mass of CME image to associate the audio and haptic feedback triggered to frequencies that will be used to calculate the known masses to verify precision of our haptic results.

CME Lesson:

As it is known the intensity of the light received on any kind of detector telescope will depend on the path followed by the light, the local conditions of the measuring device and the angle of incidence ( also the position of the device gathering the light). Being aware that currently Coronal Mass Ejections ( CME) mass determination is based on the position of the observer, the difficulty on knowing the longitudinal angle at which the CME propagated, lack of access to coronal background measurements based on results by M. A. M. Al OBAİD[16] and Aswan ( 2009)[17] we estimated that by scanning a white light coronograph image during a coronal mass ejection we could perceptually use Thomson scattering as reported by M. A. M. Al OBAİD[16] to approximate the mass of a coronal mass ejection or CME. The experience ahead cannot be taken as perception experiment or a high granularity designed test for effectiveness. The following should be taken solely as a proposal only tested on one totally congenitally blind user 25-35 of age. The user pegged the image over a piece of light. This is achieved by using a cardboard box, uniformly black in the interior and placing inside a portable flood light. A luxometer was used to verify light intensity was uniform in all areas of the portable floodlight. The portable floodlight was placed inside the box where the CME image was extended on from edge to edge of the cardboard. The CME image was marked with braille numbers at the edge of the paper. Using the orchestar to haptic the user detected and mapped the high and low pitches to map the shape of high and low intensity areas of the CME and inside the shape. By multiplying the pixels by the area and the user was able to calculate Bobs or one of the parameters needed for Thomson Scattering or Bobs [16].

Guiding our exploration by CME mass measurements performed by Aschwanden [17], the user used the sound and haptic device to map, integrate and calculate the felt and heard areas of brighter and dimmer light intensity then substitute the values calculated on the equations to approximate the mass of known X class coronal mass ejections occurred on April 7 1997 taken by LASCO and GOES and on June 20, 2013 by STEREO., at 11:24 p.m. Images were available at https://sdo.gsfc.nasa.gov/assets/gallery/preview/C2_A-IA304_1week_Nov2011_3.jpg and https://stereo-ssc.nascom.nasa.gov/browse/2013/06/20/ https://soho.nascom.nasa.gov/gallery/images/large/c24panelApr97.jpg. The result were obtained in units of $10^{15}$ g and changed to solar masses. This does not indicate our proposal is effective. Finer and higher granularity are mandated to be performed with a larger number of CME's. The masses calculated were in the range of 2.46 and 8.30 for April 7 and June 20 respectively. Given the measurements reported by Ashwaden[17] and considering the lack of correction for human error, angles, no accesss to coronal background and others we estimate this numbers could if those corrections are made , get closer to the numbers calculated by Ashwaden[17].

The authors need to corroborate with solar physicists the masses obtained and correct for existent errors and possible omissions. This results only report that it is possible to carry the experience and mainstream experiences. The experience presented herein no way indicates that haptic-sound approach is effective for calculations of masses of coronal mass ejections. For those purposes carefully well designed experiments have to happen.

5. Conclusions

In this paper we have evidenced the connection stability of a simple, low-cost device to elicit Bluetooth response of a haptic device to light intensities. The paper also reports main stream research usages of haptic, sound. It has to be tested on images and real calculations. We sure hope this simple-complex work will bring forward once again the imperative use of multisensoriality for data analysis in science.

We have presented the possible application of our methods to calculate the mass of Coronal mass ejections. Possibly a new simpler way to take into account the missing mass when using traditional methods for the determination of those masses.

Author Contributions: Wanda Diaz-Merced: discussion, conclusions, materials and methods; Ruoning Lan: introduction, materials and methods, result.

Funding: This research received no external funding. Institutional Review Board Statement: Not applicable

Acknowledgments: In loving memory of Dr. Stavros Katsanevas, for his always keen interest and enormous support in multisensorial data analysis, and more.

Conflicts of Interest: The authors declare no conflict of interest.